\documentclass[prb,amsmath,amssymb,twocolumn]{revtex4}
\usepackage{amssymb,amsbsy,bbold,mathtools,relsize,bbold,mathrsfs}
\usepackage{amsthm}
\usepackage[bbgreekl]{mathbbol}

\pdfoutput=1
\allowdisplaybreaks

\begin{document}
\title{Variational principle for magnetisation dynamics in a temperature gradient}
\author{Sylvain D. \surname{Brechet}}
\email{sylvain.brechet@epfl.ch}
\author{Jean-Philippe \surname{Ansermet}}
\affiliation{Institute of Condensed Matter Physics, Station 3, Ecole Polytechnique F\'ed\'erale de Lausanne - EPFL, CH-1015 Lausanne,
Switzerland}

\begin{abstract}

By applying a variational principle on a magnetic system within the framework of extended irreversible thermodynamics, we find that the presence of a temperature gradient in a ferromagnet leads to a generalisation of the Landau-Lifshitz equation with an additional magnetic induction field proportional to the temperature gradient. This field modulates the damping of the magnetic excitation. It can increase or decrease the damping, depending on the orientation of the magnetisation wave-vector with respect to the temperature gradient. This variational approach confirms the existence of the Magnetic Seebeck effect which was derived from thermodynamics and provides a quantitative estimate of the strength of this effect.

\end{abstract}

\maketitle

\section{Introduction}

The effect of a thermal spin torque on the magnetisation dynamics has attracted a lot of attention recently~\cite{Yu:2010, DaSilva:2012, Schellekens:2014, Choi:2015, Pushp:2015}. In a conductor, the spin dependence of the transport properties implies that a heat current induces a spin current, and consequently, a torque on the magnetisation.~\cite{Slachter:2011, Brechet:2011} In an insulator, this transport model does not apply. 

In this publication, we show that extended irreversible thermodynamics leads to a variational principle for the magnetisation which predicts the existence of an additional magnetic induction field proportional to the temperature gradient in the Landau-Lifshitz equation. In our previous work~\cite{Brechet:2013}, we called this effect the ``Magnetic Seebeck effect'' since the Seebeck effect refers to the presence of an electric field induced by a temperature gradient. This effect should not be confused with the transport phenomenon known as the spin Seebeck effect~\cite{Uchida:2008, Jaworski:2010, Uchida:2010}.

Classical irreversible thermodynamics (CIT)~\cite{Prigogine:1961, deGroot:1984, Mueller:1985} requires the system of interest to be at local equilibrium. Transport phenomena are then described by phenomenological relationships between current densities and generalised forces so as to fulfill the second law of thermodynamics. When a system does not satisfy the condition of local equilibrium, it can be described within the framework of extended irreversible thermodynamics (EIT) where the current densities are considered as additional state variables~\cite{Jou:2010}. In this article, we show that this approach provides an expression for the Magnetic Seebeck effect in terms of the thermal properties of the magnetisation.

\section{Variation of the internal energy}
In the absence of a magnetic excitation field, the magnetisation $\boldsymbol{M}$ is collinear to the magnetic induction field obtained by performing the variation of the internal energy with respect to the magnetisation $\delta u/\delta\boldsymbol{M}$, as pointed out by Gurevich.~\cite{Gurevich:1996} In the presence of a magnetic excitation field $\boldsymbol{b}$, the Landau-Lifshitz equation describes the precession of the magnetisation $\boldsymbol{M}$ about this magnetic induction field. Since the magnetisation is locally out of equilibrium, we use the framework of extended irreversible thermodynamics. According to this framework, the internal energy density $u\left(\boldsymbol{M},\boldsymbol{\nabla}\times\boldsymbol{M}\right)$ is a function of the magnetisation $\boldsymbol{M}$ and of the magnetisation current $\boldsymbol{j}_{\boldsymbol{M}} = \boldsymbol{\nabla}\times\boldsymbol{M}$ that are in turn functions of the position $\boldsymbol{r}$. According to the variational equation~\eqref{variation bis} established explicitly in the Appendix, the variational derivative of the internal energy in the bulk of the system reads,
\begin{equation}\label{Euler-Lagrange}
\frac{\delta u}{\delta\boldsymbol{M}} = \frac{\partial u}{\partial\boldsymbol{M}} + \boldsymbol{\boldsymbol{\nabla}}\times\left(\frac{\partial u}{\partial\left(\boldsymbol{\nabla}\times\boldsymbol{M}\right)}\right)
\end{equation}
The variational principle used by Gurevich~\textit{et al.}~\cite{Gurevich:1996} and Bose~\textit{et al.}~\cite{Bose:2011} assumes that the internal energy density is a function of the magnetisation $\boldsymbol{M}$ and the gradient of the magnetisation $\boldsymbol{\nabla}\,\boldsymbol{M}$. The choice of the curl of the magnetisation $\boldsymbol{\nabla}\times\boldsymbol{M}$ as a degree of the freedom of the internal energy density is motivated by the framework of extended irreversible thermodynamics. Furthermore, for a system where the magnetisation is driven out of local equilibrium, the internal energy density is expressed in the bulk as,~\cite{Stratton:1941}
\begin{equation}\label{internal energy}
u\left(\boldsymbol{M},\boldsymbol{\nabla}\times\boldsymbol{M}\right) = \frac{1}{2}\,\left(\boldsymbol{\nabla}\times\boldsymbol{M}\right)\cdot\boldsymbol{A}\left(\boldsymbol{M}\right)
\end{equation}
where $\boldsymbol{A}\left(\boldsymbol{M}\right)$ is the potential vector. This implies that the second term on the RHS of the variational equation~\eqref{Euler-Lagrange} corresponds to a magnetic induction field, as it should.

We consider a slab subjected to a temperature gradient $\boldsymbol{\nabla}\,T$ and a uniform and constant magnetic induction field $\boldsymbol{B}_0$ that are applied in plane along the $\boldsymbol{\hat{z}}$-axis as illustrated in Fig.~\ref{cone}. The magnetic excitation field $\boldsymbol{b}$ is applied orthogonally to the $\boldsymbol{\hat{z}}$-axis. The constant applied magnetic induction field $\boldsymbol{B}_0$ and the magnetic excitation field $\boldsymbol{b}$ are oriented as follows,
\begin{equation}\label{magnetic fields}
\begin{split}
&\boldsymbol{B}_0\cdot\boldsymbol{\hat{z}} = B_0\\
&\boldsymbol{b}\cdot\boldsymbol{B}_0 = 0
\end{split}
\end{equation}

The magnetisation $\boldsymbol{M}$ is the sum of the saturation magnetisation $\boldsymbol{M}_S$ and of the magnetic response field $\boldsymbol{m}$, i.e.
\begin{equation}
\boldsymbol{M} = \boldsymbol{M}_S + \boldsymbol{m}
\end{equation}
where the norms satisfy the relations
\begin{equation}\label{cond M}
\begin{split}
&\Vert\boldsymbol{M}_S\Vert \gg \Vert\boldsymbol{m}\Vert\\
&\Vert\boldsymbol{M}\Vert = M_S^2 + m^2 = \text{cste}
\end{split}
\end{equation}
and their orientations are given by,
\begin{equation}\label{magnetisations}
\begin{split}
&\boldsymbol{M}_S\cdot\boldsymbol{\hat{z}} = M_S\\
&\boldsymbol{m}\cdot\boldsymbol{M}_S = 0
\end{split}
\end{equation}

The term on the LHS and the first term on the RHS of equation~\eqref{Euler-Lagrange} are recast as,
\begin{equation}\label{def}
\begin{split}
&\frac{\delta u}{\delta\boldsymbol{M}} = \boldsymbol{B}_0 + \boldsymbol{B}_1\\
&\frac{\partial u}{\partial\boldsymbol{M}} = \boldsymbol{B}_0 + \boldsymbol{b}
\end{split}
\end{equation}
where the first-order magnetic induction field $\boldsymbol{B}_1$ is oriented as follows,
\begin{equation}
\boldsymbol{B}_1\cdot\boldsymbol{B}_0 = 0
\end{equation}
The temperature gradient $\boldsymbol{\nabla}\,T$ is imposed along the $\boldsymbol{\hat{z}}$-axis as illustrated in Fig.~\ref{cone}. Along that axis, the spatial symmetry of the system is broken, i.e.
\begin{equation}\label{broken}
\boldsymbol{\nabla} = \boldsymbol{\hat{z}}\,\frac{\partial}{\partial z}
\end{equation}
The conditions~\eqref{magnetisations} and~\eqref{broken} imply that,
\begin{equation}\label{curl magnetisation}
\boldsymbol{\nabla}\times\boldsymbol{M} = \boldsymbol{\nabla}\times\boldsymbol{m}
\end{equation}
The relations~\eqref{def} and~\eqref{curl magnetisation} imply that the variational equation~\eqref{Euler-Lagrange} is recast as,
\begin{equation}\label{Euler-Lagrange bis}
\boldsymbol{B}_1 = \boldsymbol{b} + \boldsymbol{\boldsymbol{\nabla}}\times\left(\frac{\partial u}{\partial\left(\boldsymbol{\nabla}\times\boldsymbol{m}\right)}\right)
\end{equation}

\section{Magnetic Seebeck effect}
The second term on the RHS of the relation~\eqref{Euler-Lagrange bis} is recast as,
\begin{equation}\label{2nd term}
\boldsymbol{\nabla}\times\left(\frac{\partial u}{\partial\left(\boldsymbol{\nabla}\times\boldsymbol{m}\right)}\right) = \left(\boldsymbol{\nabla}^{-1}\times\left(\frac{\partial\left(\boldsymbol{\nabla}\times\boldsymbol{m}\right)}{\partial u}\right)\right)^{-1}
\end{equation}
In the linear response, i.e. to first-order in the magnetic excitation field $\boldsymbol{m}$, the differential operator $\partial/\partial u$ commutes with the differential operator $\boldsymbol{\nabla}$ which implies that,
\begin{equation}\label{2nd term bis}
\frac{\partial\left(\boldsymbol{\nabla}\times\boldsymbol{m}\right)}{\partial u} = \boldsymbol{\nabla}\times\frac{\partial\boldsymbol{m}}{\partial u} 
\end{equation}
Substituting the relation~\eqref{2nd term bis} into the expression~\eqref{2nd term} yields,
\begin{equation}\label{2nd term ter}
\boldsymbol{\nabla}\times\left(\frac{\partial u}{\partial\left(\boldsymbol{\nabla}\times\boldsymbol{m}\right)}\right) = \boldsymbol{\nabla}\times\left(\boldsymbol{\nabla}^{-1}\times\frac{\partial u}{\partial\boldsymbol{m}}\right)
\end{equation}
In the linear response, the relations~\eqref{magnetic fields}-\eqref{def} imply that,
\begin{equation}\label{deriv m}
\frac{\partial u}{\partial\boldsymbol{m}} = \boldsymbol{b}
\end{equation}
Taking into account the conditions~\eqref{broken},~\eqref{deriv m} and
\begin{equation}\label{nabla id}
\boldsymbol{\nabla}\cdot\boldsymbol{\nabla}^{-1} = 1
\end{equation}
the expression~\eqref{2nd term ter} becomes,
\begin{equation}\label{2nd term quad}
\boldsymbol{\nabla}\times\left(\frac{\partial u}{\partial\left(\boldsymbol{\nabla}\times\boldsymbol{m}\right)}\right) = -\,\boldsymbol{\hat{z}}\cdot\boldsymbol{\nabla}^{-1}\left(\frac{\partial}{\partial z}\,\boldsymbol{b}\right)
\end{equation}
which implies that the relation~\eqref{Euler-Lagrange bis} is recast as,
\begin{equation}\label{Euler-Lagrange bis 2}
\boldsymbol{B}_1 = \boldsymbol{b} -\,\boldsymbol{\hat{z}}\cdot\boldsymbol{\nabla}^{-1}\left(\frac{\partial}{\partial z}\,\boldsymbol{b}\right)
\end{equation}
The linear magnetic constitutive equation reads,
\begin{equation}\label{mag const equation}
\boldsymbol{b} = \mu_0\,\chi^{-1}\,\boldsymbol{m}
\end{equation}
which implies that the relation~\eqref{Euler-Lagrange bis 2} becomes,
\begin{equation}\label{Euler-Lagrange hep}
\boldsymbol{B}_{1} = \boldsymbol{b} -\,\mu_0\,\boldsymbol{\hat{z}}\cdot\boldsymbol{\nabla}^{-1}\left(\frac{\partial}{\partial z}\left(\chi^{-1}\,\boldsymbol{m}\right)\right)
\end{equation}
The derivative of the second condition~\eqref{cond M} yields,
\begin{equation}\label{cond M bis}
M_S\,\frac{\partial M_S}{\partial z} + m\,\frac{\partial m}{\partial z} = 0
\end{equation}
which implies that for a uniform precession in the plane orthogonal to the $\boldsymbol{\hat{z}}$-axis,
\begin{equation}\label{cond M ter}
\frac{\partial}{\partial z}\,\boldsymbol{m} = -\,\frac{M_S}{m^2}\left(\frac{\partial M_S}{\partial z}\right)\boldsymbol{m} 
\end{equation}
The saturation magnetisation $M_S = M_S\left(T\right)$ is a function of the temperature, which implies that,
\begin{equation}\label{temp dep}
\frac{\partial M_S}{\partial z} = \frac{\partial M_S}{\partial T}\left(\boldsymbol{\hat{z}}\cdot\boldsymbol{\nabla}\,T\right)
\end{equation}
The magnetic susceptibility $\chi$ is a function of the saturation magnetisation $M_S$, i.e.
\begin{equation}\label{temp dep 2}
\frac{\partial\,\chi}{\partial z} = \frac{\partial\,\chi}{\partial M_S}\,\frac{\partial M_S}{\partial T}\left(\boldsymbol{\hat{z}}\cdot\boldsymbol{\nabla}\,T\right)
\end{equation}

\begin{figure}[htb]
\begin{center}
\includegraphics[width=\columnwidth]{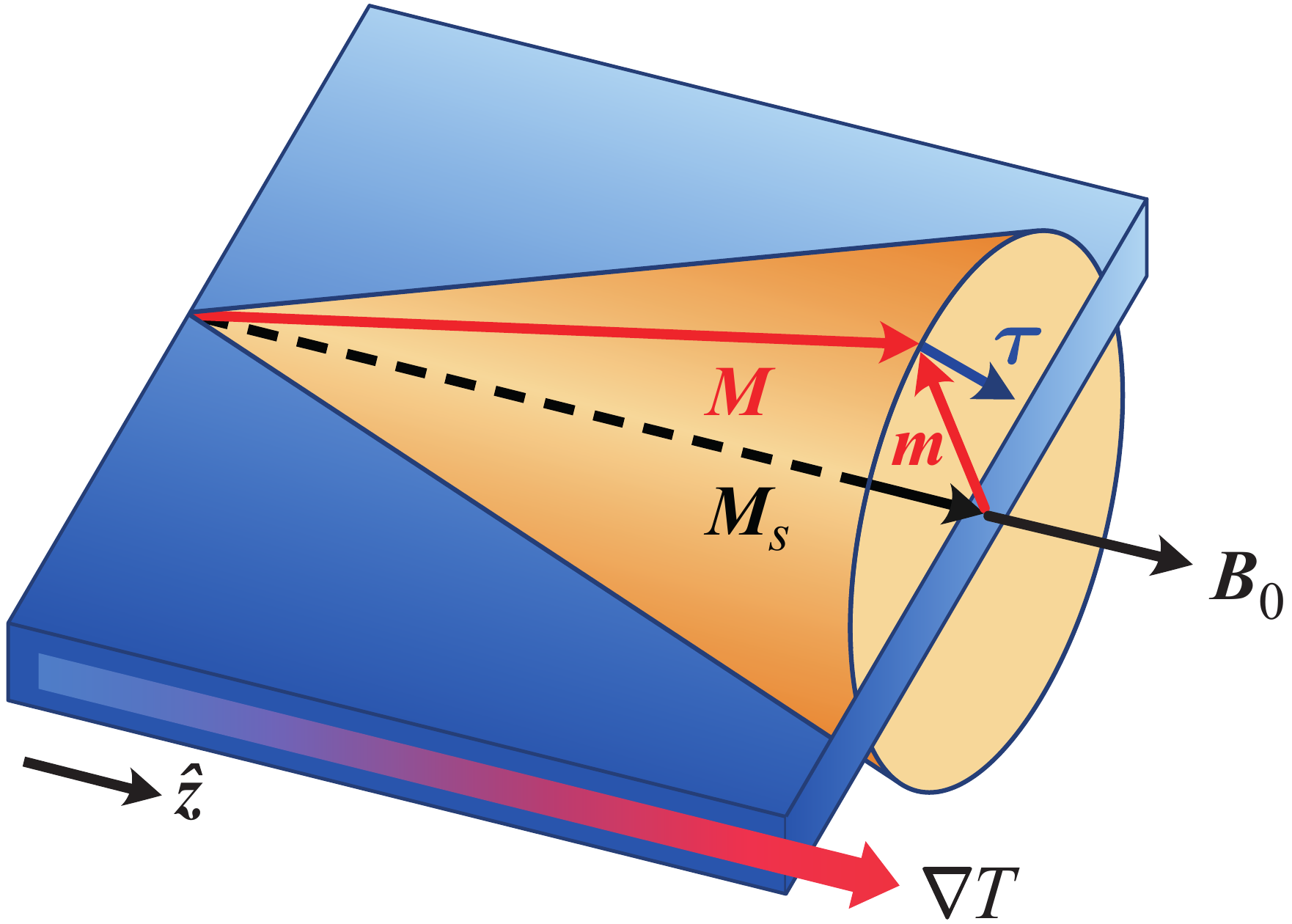}
\caption{Precession cone of the magnetisation $\boldsymbol{M}$ where the saturation magnetisation $\boldsymbol{M}_S$, the constant magnetic induction field $\boldsymbol{B}_0$ and the temperature gradient $\boldsymbol{\nabla}\,T$ are oriented in plane along the $\boldsymbol{\hat{z}}$-axis. The magnetic response field $\boldsymbol{m}$ and the magnetic torque $\boldsymbol{\tau}$ are orthogonal to the $\boldsymbol{\hat{z}}$-axis.}
\label{cone}
\end{center}
\end{figure}

In the linear response, the relations~\eqref{cond M ter}-\eqref{temp dep 2} imply that  the expression~\eqref{Euler-Lagrange hep}
is recast as,
\begin{equation}\label{Euler-Lagrange oct 0}
\boldsymbol{B}_{1} = \boldsymbol{b} + \frac{\mu_0}{\chi}\!\left(\frac{1}{\chi}\,\frac{\partial\,\chi}{\partial M_S} + \frac{M_S}{m^2}\right)\!\left(\frac{\partial M_S}{\partial T}\right)\left(\boldsymbol{\nabla}\,T\right)\cdot\boldsymbol{\nabla}^{-1}\boldsymbol{m}
\end{equation}
For a ferromagnet, the magnetic susceptibility $\chi$ is proportional to the saturation magnetisation $M_S$~\cite{Morrish:2001}. Thus, the first condition~\eqref{cond M} imposes on the first term in brackets~\eqref{Euler-Lagrange oct 0} the condition,
\begin{equation}\label{cond extra}
\frac{1}{\chi}\,\frac{\partial\,\chi}{\partial M_S} = \frac{1}{M_S} \ll \frac{M_S}{m^2}
\end{equation}
Taking into account the condition~\eqref{cond extra}, the relation~\eqref{Euler-Lagrange oct 0} reduces to,
\begin{equation}\label{Euler-Lagrange oct}
\boldsymbol{B}_{1} = \boldsymbol{b} + \frac{\mu_0\,M_S}{\chi\,m^2}\left(\frac{\partial M_S}{\partial T}\right)\left(\boldsymbol{\nabla}\,T\right)\cdot\boldsymbol{\nabla}^{-1}\boldsymbol{m}
\end{equation}
The first-order magnetic induction field $\boldsymbol{B}_{1}$ that is orthogonal to the zeroth-order magnetisation saturation $\boldsymbol{M}_S$ exerts a thermal magnetic torque $\boldsymbol{\tau} = \boldsymbol{M}_S\times\boldsymbol{B}_{1}$~\cite{Brechet:2013b} on the magnetisation $\boldsymbol{M}$ illustrated in Fig.~\ref{cone}.

To compare the result of this analysis with our previous work on the Magnetic Seebeck effect~\cite{Brechet:2013}, we recast the relation~\eqref{Euler-Lagrange oct} as,
\begin{equation}\label{Euler-Lagrange nov}
\boldsymbol{B}_{1} = \boldsymbol{b} -\,\mu_0\left(\boldsymbol{k}_T\cdot\boldsymbol{\nabla}^{-1}\right)\boldsymbol{m}
\end{equation}
where the thermal wave-vector $\boldsymbol{k}_T$ is given by,
\begin{equation}\label{thermal wave vector}
\boldsymbol{k}_T = -\,\frac{M_S}{\chi\,m^2}\left(\frac{\partial M_S}{\partial T}\right)\,\boldsymbol{\nabla}\,T
\end{equation}
where $\partial M_S/\partial T <0$. The expression~\eqref{Euler-Lagrange nov} for the magnetic induction field $\boldsymbol{B}_{1}$ has the same structure as equation $(7)$ of our previous work~\cite{Brechet:2013}. However, in the second term on the RHS accounting for the Magnetic Seebeck effect, the expression of the thermal wave-vector $\boldsymbol{k}_T$ differs. According to our previous work in the classical irreversible thermodynamical framework~\cite{Brechet:2013}, the expression for the thermal wave-vector $\boldsymbol{k}_T$ is,
\begin{equation}\label{thermal wave vector bis}
\boldsymbol{k}_T = \frac{\lambda\,n\,k_B}{\mu_0\,M_S^2}\,\boldsymbol{\nabla}\,T
\end{equation}
where $n$ is the Bohr magneton number density, $k_B$ is Boltzmann's constant and $\lambda$ is the dimensionless Magnetic Seebeck parameter. Identifying the relations~\eqref{thermal wave vector} and~\eqref{thermal wave vector bis} yields the following expression for the Magnetic Seebeck parameter~\footnote{This expression yields an estimate for the Magnetic Seebeck parameter $\lambda = 7\cdot 10^{-5}$, using the values of the parameters cited in reference~\cite{Brechet:2013}. It is of the same order of magnitude as the rough estimate based on the observations of reference~\cite{Brechet:2013}.},
\begin{equation}\label{MS para}
\lambda = -\,\frac{\mu_0\,M_S^2}{n\,k_B}\left(\frac{m}{M_S}\right)^{-2}\!\left(\frac{1}{\chi\,M_S}\,\frac{\partial M_S}{\partial T}\right) > 0
\end{equation}
since $\partial M_S/\partial T <0$. The thermal vector $\boldsymbol{k}_T$ generates a magnetic induction field~\eqref{Euler-Lagrange nov} that is proportional to the temperature gradient $\boldsymbol{\nabla}\,T$. This field leads to a generalisation of the Landau-Lifshitz equation. The respective orientation between the wave-vector $\boldsymbol{k}$ of the magnetisation waves and the temperature gradient $\boldsymbol{\nabla}\,T$ leads to an increase or attenuation of the magnetic damping. As predicted theoretically and observed experimentally for magnetostatic backward volume modes in a YIG slab~\cite{Brechet:2013}, the magnetic damping is attenuated for magnetisation waves propagating along the temperature gradient while it is increased for magnetisation waves propagating against the temperature gradient. This modulation of the magnetic damping is a consequence of the Magnetic Seebeck effect.

\section{Conclusion}
In this article, the variational approach for the description of magnetisation dynamics is performed in the extended irreversible thermodynamical framework. It predicts the existence of a Magnetic Seebeck effect for the propagation of magnetisation waves along a temperature gradient in a magnetic slab. In the classical irreversible thermodynamical  framework, the coupling between the magnetisation dynamics and the thermal gradient is expressed by a phenomenological relation imposed in order to satisfy the second principle of thermodynamics. By contrast, in the extended irreversible thermodynamical framework, the coupling between the magnetisation dynamics and the thermal gradient is intrinsic to the description of the system itself, i.e. it derives from the thermal properties of the system. The comparison between the expressions obtained for the thermal wave-vector $\boldsymbol{k}_T$ in both frameworks yields an explicit expression for the dimensionless parameter $\lambda$ that defines the strength of the Magnetic Seebeck effect.

\appendix

\section{Variational principle}\label{Variational derivative}

In a stationary state, the kinetic energy associated with the precession is constant. Thus, it can be ignored while performing the action variation. The action $S\left(\boldsymbol{M},\boldsymbol{\nabla}\times\boldsymbol{M}\right)$ and the Lagrangian density $\mathcal{L}\left(\boldsymbol{M},\boldsymbol{\nabla}\times\boldsymbol{M}\right)$ are functions of the magnetisation $\boldsymbol{M}$ and of the magnetisation current $\boldsymbol{\nabla}\times\boldsymbol{M}$ that are in turn functions of the position $\boldsymbol{r}$. These quantities are related by the integral expression,
\begin{equation}\label{action}
S\left(\boldsymbol{M},\boldsymbol{\nabla}\times\boldsymbol{M}\right) = \int dt\,d^3r\,\mathcal{L}\left(\boldsymbol{M},\boldsymbol{\nabla}\times\boldsymbol{M}\right)
\end{equation}
In a stationary state, the internal energy density $u\left(\boldsymbol{M},\boldsymbol{\nabla}\times\boldsymbol{M}\right)$ is equal to the opposite of the Lagrangian density $\mathcal{L}\left(\boldsymbol{M},\boldsymbol{\nabla}\times\boldsymbol{M}\right)$ up to a constant, i.e.
\begin{equation}\label{L-u}
\mathcal{L}\left(\boldsymbol{M},\boldsymbol{\nabla}\times\boldsymbol{M}\right) = -\,u\left(\boldsymbol{M},\boldsymbol{\nabla}\times\boldsymbol{M}\right)
\end{equation}
where the internal energy density plays the role of the potential. Thus, the action~\eqref{action} is recast as, 
\begin{equation}\label{action bis}
S\left(\boldsymbol{M},\boldsymbol{\nabla}\times\boldsymbol{M}\right) = -\int dt\,d^3r\,u\left(\boldsymbol{M},\boldsymbol{\nabla}\times\boldsymbol{M}\right)
\end{equation}
The variation of the action $S\left(\boldsymbol{M},\boldsymbol{\nabla}\times\boldsymbol{M}\right)$ is expressed formally as,
\begin{equation}\label{action variation}
\delta S = -\int dt\,d^3r\,\delta u
\end{equation}
and yields,
\begin{equation}\label{action variation bis}
\delta S = -\int dt\,d^3r\left[\,\frac{\partial u}{\partial\boldsymbol{M}}\cdot\delta\boldsymbol{M} + \frac{\partial u}{\partial\left(\boldsymbol{\nabla}\times\boldsymbol{M}\right)}\cdot\delta\left(\boldsymbol{\nabla}\times\boldsymbol{M}\right)\right]
\end{equation}
Using the vectorial identity,
\begin{equation}\label{vect id}
\boldsymbol{A}\cdot\delta\left(\boldsymbol{\nabla}\times\boldsymbol{M}\right) = \left(\boldsymbol{\nabla}\times\boldsymbol{A}\right)\cdot\delta\boldsymbol{M} -\,\boldsymbol{\nabla}\cdot\left(\boldsymbol{A}\times\delta\boldsymbol{M}\right) 
\end{equation}
the relation~\eqref{action variation bis} is recast as,
\begin{align}\label{action variation ter}
&\delta S = -\int dt\,d^3r\left[\,\frac{\partial u}{\partial\boldsymbol{M}} + \boldsymbol{\nabla}\times\left(\frac{\partial u}{\partial\left(\boldsymbol{\nabla}\times\boldsymbol{M}\right)}\right)\,\right]\cdot\delta\boldsymbol{M}\nonumber\\*
&\phantom{\delta S =} + \int dt\,d^3r\,\boldsymbol{\nabla}\cdot\left(\frac{\partial u}{\partial\left(\boldsymbol{\nabla}\times\boldsymbol{M}\right)}\times\delta\boldsymbol{M}\right) 
\end{align}
Using the divergence theorem, the second integral on the RHS of the relation~\eqref{action variation ter} is recast as a surface integral. Taking the limit where the integration surface tends to infinity and assuming that the magnetisation is uniform at infinity, this integral vanishes, which implies that the relation~\eqref{action variation ter} reduces to,
\begin{equation}\label{action variation quad}
\delta S = -\int dt\,d^3r\left[\,\frac{\partial u}{\partial\boldsymbol{M}} + \boldsymbol{\nabla}\times\left(\frac{\partial u}{\partial\left(\boldsymbol{\nabla}\times\boldsymbol{M}\right)}\right)\,\right]\cdot\delta\boldsymbol{M}
\end{equation}
Identifying the integrands in relations~\eqref{action variation} and~\eqref{action variation quad} and taking the variational derivative of the functional $u\left(\boldsymbol{M},\boldsymbol{\nabla}\times\boldsymbol{M}\right)$ with respect to $\boldsymbol{M}$, also known as functional derivative~\cite{Mathews:1970} yields,
\begin{equation}\label{variation bis}
\frac{\delta u}{\delta\boldsymbol{M}} = \frac{\partial u}{\partial\boldsymbol{M}} + \boldsymbol{\boldsymbol{\nabla}}\times\left(\frac{\partial u}{\partial\left(\boldsymbol{\nabla}\times\boldsymbol{M}\right)}\right)
\end{equation}
\bibliography{references}
\end{document}